# 1T-FeS$_2$: a new type of two-dimensional metallic ferromagnet

Govindan Kutty Rajendran Nair[1†], Xiaoyu Ji[2,3†], Dong Guo[2†], Chao Zhu[1†], Xiaodong Xu[4], Xinyi Zheng[2,5], Xue Yang[2,5], Jian Cui[6], Peiling Li[2], Xiaowei Wang[1], Wu Yao[1], Jiadong Zhou[7], Teddy Salim[1], Jian Yi[8], Fengcai Ma[3], Changli Yang[2], Hua Ke[4], Fanming Qu[2,5,9], Jie Shen[2,5,9], Xiunian Jing[2,9], Zheng Liu[1,10,11*], Xingji Li[4,*], Guangtong Liu[2,5,9*], and Li Lu[2,5,9]

[1]School of Materials Science and Engineering, Nanyang Technological University, 50 Nanyang Avenue, 639798, Singapore

[2]Beijing National Laboratory for Condensed Matter Physics, Institute of Physics, Chinese Academy of Sciences, Beijing 100190, China

[3]Department of Physics, Liaoning University, Shenyang 110036, China

[4]School of Material Science and Engineering, Harbin Institute of Technology, Harbin 100150, China

[5]School of Physical Sciences, University of Chinese Academy of Sciences, Beijing 100049, China

[6]Beijing Academy of Quantum Information Sciences, Beijing 100193, China

[7] School of Physics, Beijing Institute of Technology, Beijing 100081, China

[8]Ningbo Institute of Industrial Technology, Chinese Academy of Sciences, 315201 Ningbo, China.

[9]Songshan Lake Materials Laboratory, Dongguan, Guangdong 523808, China

[10]CINTRA CNRS/NTU/THALES, UMI 3288, Research Techno Plaza, Singapore 637553, Singapore

[11]School of Electrical and Electronic Engineering, Nanyang Technological University, Singapore 639798, Singapore

†These authors contributed equally to this work. Correspondence and requests for materials should be addressed to ZL (email: z.liu@ntu.edu.sg), XL (email: lxj0218@hit.edu.cn) and GL (email: gtliu@iphy.ac.cn)

**Abstract**

Discovery of intrinsic two-dimensional (2D) magnetic materials is crucial for understanding the fundamentals of 2D magnetism and realizing next-generation magnetoelectronic and magneto-optical devices. Although significant efforts have been devoted to identifying 2D magnetism by exfoliating bulk magnetic layered materials, seldom studies are performed to synthesize ultra-thin magnetic materials directly for non-layered magnetic materials. Here, we report the successful synthesis of a new type of theoretically proposed 2D metallic ferromagnet—1T-FeS$_2$, through the molten-salt-assisted chemical vapor deposition (CVD) method. The long-range 2D ferromagnetic order is confirmed by the observation of a large anomalous Hall effect (AHE) and a hysteretic magnetoresistance. The experimentally detected out-of-plane ferromagnetic ordering is theoretically suported with Stoner criterion *ST* > 1. Our findings open up new possibilities to search novel 2D ferromagnets in non-layered compounds and render opportunities for realizing realistic ultra-thin spintronic devices.

**Introduction**

Two-dimensional (2D) metallic magnets have attracted intense attention because of their rich physics and potential applications in nanoscale spintronic devices.[1, 2] Though magnetic order is theoretically prohibited in isotropic 2D systems at finite temperatures according to Mermin–Wagner theorem, the presence of magnetic anisotropy can counteract the thermal fluctuations and establish the long-range magnetic order.[3] The significant advances in the field of 2D magnet include the recently discovered 2D ferromagnetism (FM) in monolayer $CrI_3$[4], bilayer $Cr_2Ge_2Te_6$[5], multilayer $CrTe_2$[6], $VSe_2$[7], $CrTe_2$[8], $1T-MnSe_2$[9], $Fe_3GeTe_2$[10], etc. To date, significant efforts have been devoted to identifying 2D FM by thinning down magnetic layered bulk materials with mechanical exfoliation method.[11, 12, 13, 14] The poor layer-control, low yield and small size (several micrometers) of exfoliated thin flakes greatly limit the fundamental studies and practical applications of those 2D magnets in spintronics. Chemical vapor deposition (CVD) can offer an alternate pathway to synthesize large-area, scalable, layer-controllable, cost-effective and high-quality 2D magnets. Only a few reports are available on the CVD synthesis of 2D magnets.[15, 16, 17] Thus it is urgent and quintessential to explore the feasibility of synthesizing more 2D magnets via CVD. In addition, it is fundamentally important to search for more 2D materials in layered/non-layered materials from a scientific point of view. Recently, great progress in the field of non-layered 2D-like magnets includes the successful isolation of single-layer $ZnSe$[18], $Fe_2O_3$[19], and $FeS_2$[20] from their non-layered bulk counterparts. More importantly, the FM observed in 2D $Fe_2O_3$ highlights the influence of quantum confinement on the electronic structure, which enlightens us to explore 2D intrinsic FM in $1T-FeS_2$, a new form of another kind of natural iron ore Pyrite ($β-FeS_2$).

Iron Pyrite ($\beta$-FeS$_2$) is a low-cost, non-toxic, and earth-abundant semiconductor material. The suitable bandgap (0.95 eV) and high optical absorption coefficient (>10$^5$ cm$^{-1}$) make FeS$_2$ great potential applications in green energy fields.[21, 22, 23, 24, 25, 26] $\beta$-FeS$_2$ was known as non-ferromagnetic because of the low spin state of the iron atom (Fe$^{2+}$).[27, 28, 29] Interestingly, a new kind of 2D FeS$_2$, including T-FeS$_2$ and S-FeS$_2$ monolayers were recently predicted to exist at room temperature.[30] More importantly, T-FeS$_2$ and S-FeS$_2$ were predicted to possess metallic ferromagnetism and semiconducting antiferromagnetism, emphasizing the quantum confinement effect on the physical properties of a material. Although FeS$_2$ nanocrystals have been recently synthesized, the synthesis of atomically thin T-FeS$_2$ and S-FeS$_2$ flakes at present remains experimentally challenging. Furthermore, to investigate the theoretically proposed novel magnetic properties, it is fundamentally important to prepare 2D FeS$_2$ at first.

In this work, we report the successful synthesis of 1T-FeS$_2$ ferromagnetic metal layers via the chemical vapor deposition (CVD) method. The atomic structure of 1T-FeS$_2$ has been characterized by complementary tools, including Raman spectra and scanning transmission electron microscopy (STEM). Low-temperature magnetotransport measurements show that the as-prepared atomically thin 1T-FeS$_2$ flake is a two-dimensional (2D) ferromagnetic metal with strong magnetic anisotropy, evidenced by a large anomalous Hall effect and first-principles calculations. These results establish 1T-FeS$_2$ ultrathin films as a promising 2D ferromagnet for exotic low-dimensional spintronics applications.

**Results**

**Growth of 2D 1T-FeS$_2$ crystals.** 2D and few-layer 1T-FeS$_2$ crystals were synthesized using a molten-salt-assisted chemical vapor deposition (CVD) method (Supplementary Fig. 1).[31] In our experiments, Iron chloride (FeCl$_2$) and sulfur (S) were selected as precursors (See the Methods for detailed information on synthesis). Figure 1a shows the crystal structure of monolayer 1T-FeS$_2$ viewed from different angles. The as-grown few-layer FeS$_2$ crystals are typically triangular with sharp edges, as can be seen from Fig. 1b,c. The uniform color and sharp edges reflect the high-quality single-crystalline and homogenous elemental distribution, as revealed by the following TEM and EDX characterizations (Fig. 2). The as-prepared few-layer triangular FeS$_2$ crystals have a lateral size up to 110 μm, which allows us to study its novel physical properties via transport measurements. Figure 1c displays a representative atomic force microscopy (AFM) measurement taken on a few-layer FeS$_2$ crystal, which has a thickness of ~2 nm, indicating it is a ~ trilayer sample. Though monolayer FeS$_2$ can be obtained experimentally (inset of Fig. 1g and Supplementary Fig. 3a), we could not get reliable AFM data due to its sensitivity to the ambient environment. We note that the ambient CVD method is a controllable synthetic strategy to prepare FeS$_2$ layers. Figures 1d-f show the distribution statistics of layer number ($L$) and representative morphologies of FeS$_2$ crystals grown at different growth temperatures ($T_g$). Obviously, the layer number of obtained FeS$_2$ crystals increases with $T_g$, agreeing with our previous observations in NbSe$_2$.[32] Monolayer FeS$_2$ crystals were mainly obtained when $T$ was set to 600–620 °C. More information about the controllable growth of FeS$_2$ layers can be found in Methods.

The chemical states of the as-grown FeS$_2$ samples were examined by X-ray photoelectron spectroscopy (XPS). Figures 1g and 1h display the 2$p$ core levels spectra for elements Fe and S, respectively. Consistent with previous studies of FeS$_2$ in bulk and nanoforms, two peaks at 707.2 eV and 719.8 eV are ascribed to Fe$^{2+}$ 2$p_{3/2}$ and Fe$^{2+}$ 2$p_{1/2}$ (Fig. 1g), and the peaks at 163.4 eV and 162.3 eV (Fig. 1h) correspond to S 2$p_{3/2}$ and S 2$p_{1/2}$.[33, 34, 35] More importantly, there are two new peaks centered at 709.5 eV and 723.8 eV compared to previous studies, which belong to Fe$^{3+}$ 2$p_{3/2}$ and $Fe^{3+}$ $2p_{1/2}$, respectively.[33, 36, 37] It is noted that $Fe^{3+}$ had never been detected in $FeS_2$ before. We believe the existence of $Fe^{3+}$ results in the ferromagnetism of 1T-$FeS_2$, as discussed below. Furthermore, the absence of O 2$p$ peaks in Supplementary Fig. 4 suggests that the as-grown $FeS_2$ few-layers are high quality. Figure 1i is a typical Raman spectrum collected on an as-grown 1T-$FeS_2$ sample (see Supplementary Fig. 2 for more data), which exhibits three distinctive peaks centered around 337.70 cm$^{-1}$, 374.71 cm$^{-1}$ and 424.07 cm$^{-1}$, respectively. These peaks are attributed to the $A_g$ (337.70 cm$^{-1}$), $E_g$ (374.71 cm$^{-1}$), and $T_g$ (424.07 cm$^{-1}$) mode, respectively. A redshift in the peak values with respect to the reported conventional Raman active peak value was observed.[38] As the layer number decreases, the intensity of these Raman active modes weakens, consistent with the previous report for other 2D materials.[32, 39]

**Structural characterization.** Atomic-resolution scanning transmission electron microscopy (STEM) imaging and energy-dispersive X-ray spectrometry (EDX) were employed to characterize the atomic structure and the chemical composition of $FeS_2$ few layers. Figure 2a shows a high-resolution Z-contrast STEM image of a triangular 1T-$FeS_2$ flake, where the octahedral atomic lattice of alternating bright and dark spots, which corresponds to the Fe and $S_2$ atomic columns, can be resolved. The high crystallinity of the as-grown 1T-$FeS_2$ is further supported by the Fast Fourier transform (FFT) pattern shown in the top right of Fig. 2a. The interplanar $d$-spacing of (100) plane is measured to be 2.91 Å, which is consistent with the reported value of N-doped $FeS_2$, while slightly larger than that of $FeS_2$ bulk phase (2.70 Å) (JCPDS 42–1340).[40] As shown in the inset of Fig. 2a, the excellent agreement between the experimental and the simulated STEM images (the bottom right of Fig. 2a) confirms that the synthesized $FeS_2$ few layers crystallize in the 1T phase.

Figure 2h shows the sulfur vacancies and Fig.2i reveals the point-defects in the 1T-FeS$_2$ lattice. Similar phenomena were found in CVD-grown NbSe$_2$ flakes.[32] The defect and vacancy concentration are estimated to be ~2.4% and 4.7%, demonstrating the sufficiently low defect concentration present in the few-layer FeS$_2$ lattice. EDS was used to identify the chemical constituents of the as-grown layers. Figure 2e,f and g display the elemental mapping results of Fe and S, which unambiguously shows that Fe and S are uniformly distributed in the crystal without the presence of any other impurities. The elemental ratio of Fe:S was estimated to be 1:1.8 from Fig. 2j, close to 1:1.90 extracted from the STEM result.

**Low-temperature magnetotransport experiments.** To further characterize the quality of as-grown few-layer 1T-FeS$_2$ flakes, we fabricated Hall-bar devices (inset of Fig. 3a and Supplementary Fig. 8) and performed low-temperature magnetotransport measurements. Figure 3a shows the temperature-dependent sheet resistance $R_{xx}(T)$ measured at zero magnetic field for five typical devices (denoted by S1~S5) fabricated from different batches of as-grown 1T-FeS$_2$. Surprisingly, we find that all the ultra-thin 1T-FeS$_2$ devices exhibit a metallic conducting behavior with $dR/dT>0$, which is in stark contrast to its semiconducting bulk form. We note that though a metallic transport property has been predicted for atomically thin T-FeS$_2$ samples, no metal conductivity is experimentally reported yet.[30] Figures 3b and 3c plot the longitudinal magnetoresistance MR [defined as $\text{MR} = \frac{R_{xx}(\mu_0 H) - R_{xx}(0)}{R_{xx}(0)} \times 100\%$] and Hall resistance $R_{xy}(H)$ collected on device S1 (see Supplementary Fig. 8 for more data). From Fig. 3b, no MR can be discernable at $T>50$ K. However, a small negative MR begins to emerge at $T = 50$ K and further evolves into a value of -5.0% at $T = 20$ K under a perpendicular magnetic field $\mu_0 H_\perp = 11$ T. Correspondingly, Hall traces remains linear until $T$ is reduced to 50 K in Fig. 3c. Then, $R_{xy}(H)$ gradually deviates from the linear behavior and transforms into an "S" shape at $T<22$ K, implying the presence of anomalous Hall effect (AHE) as demonstrated below. Thus, the carrier type and density can be extracted and summarized in Supplementary Table 1. Clearly, the carrier (electron) has a sheet density of ~$10^{16}$ cm$^{-2}$, which is in good agreement with the typical value of $10^{28}$ m$^{-3}$ of metals by considering the sample thickness of ~5 nm, corroborating the ultra-thin 1T-FeS$_2$ flake is a metal.

The few-layer FeS$_2$ crystals provide an ideal platform to study the magnetism in 2D limit. Figure 4 illustrates that an unprecedented long-range ferromagnetic order has been realized in few-layer 1T-FeS$_2$ crystals by displaying anomalous Hall resistance $R_{xy}^{A}(H)$ at $T$<20 K. Figures 4a shows $R_{xy}^{A}(H)$ observed in device S4 (see Supplementary Fig. 9 for raw data), where the admixing of $R_{xx}(H)$ and the conventional Hall effect $R_{xy}^{N}(H)$ contributions are removed (see Supplementary Information for details). Obviously, a magnetic hysteresis loop appears in $R_{xy}^{A}(H)$ traces at $T$=2 K, which is a characteristic of anomalous Hall effect.[41, 42, 43] As the temperature increases from 2 K to 15 K, the loop area shrinks gradually and finally disappears at $T$~20 K, consistent with the S-shaped Hall traces observed below 22 K in Fig. 3c. Concomitantly, the magnetic hysteresis superimposed on $R_{xx}(H)$ traces varnishes around 20 K (Supplementary Fig. 9a). These features, including the AHE, negative MR, and low-field MR hysteresis, occurred below 20 K, are hallmarks of ferromagnetic order. Figure 4c summarizes the coercive field $H_c$ as a function of temperature, where the red dashed line represents a smooth fit of $H_c(T)$ data points with an empirical formula $H_c(T) \propto (1 - T/T_c)^\beta$, where $T_C$ is the Curie temperature. The best-fitting yields $T_C$ = 15.03 K, which is quantitatively in good agreement with the experimental observation of magnetic hysteresis emerged at $T$ = 15 K, as well as the peak value of $T$ = 15.3 K in the differential resistance ($dR/dT$) curve (inset of Fig. 4c). The strongly temperature-dependent coercivity $H_c(T)$ observed here rules out the possibility of defect-induced ferromagnetism.[44]

Similar results were observed in device S5 (Supplementary Fig. 10). The data shown above demonstrate that few-layer 1T-FeS$_2$ crystal is a new type of 2D metallic ferromagnet.

Furthermore, we have conducted tilt experiments to determine the easy axis of magnetization. The corresponding AHE results $R_{xy}^A(H)$ are shown in Fig. 4b (see Supplementary Fig. 11 for raw data). Clearly, the magnetic hysteresis loop evolves into a maximum as the magnetic field is tilted from out-of-plane to in-plane configuration. A more detailed insight into the tilt angle dependence of the magnetization is shown in Fig. 4d, where the coercive field $H_C(\theta)$ versus tilt angle $\theta$ is represented. From Figure 4d, one can see that $H_C(\theta)$ remains nearly constant at 0.027 T with $\theta$ <120°. As $\theta$ is further increased, $H_C(\theta)$ increases steeply and reaches a maximum of 0.448 T at $\theta$=180° followed by a rapid decrease, confirming a large magnetic anisotropy existed in few-layer 1T-FeS$_2$. This extremely sensitive angle-dependent coercive field $H_C(\theta)$ indicates that the easy axis of magnetization is along the *c*-axis. We note that the large magnetic anisotropy is a necessary ingredient to form long-range magnetic order in the 2D limit, as recently observed in layer-dependent ferromagnetism.[4]

The emergent 2D metallic ferromagnetism in few-layer 1T-FeS$_2$ is surprising, due to its bulk counterpart being a non-ferromagnetic semiconductor. Recently, Zhang *et al* predicated two new stable phases of S-FeS$_2$ and T-FeS$_2$ that can exist in two-dimensional layered material down to the atomic monolayer limit.[30] Moreover, the monolayer T-FeS$_2$ is predicated to be a 2D metallic ferromagnet. We find these predications are perfectly consistent with our observations of STEM and magnetotransport measurements. Different from the generally observed low-spin state of $Fe^{2+}$ in bulk FeS$_2$, our XPS data clearly show that $Fe^{3+}$ exists in the high-spin state, which is supposed to be the origin of the ferromagnetism in the 1T-FeS$_2$ crystals. The assignment is also consistent with the theoretically predicted 2D ferromagnetism originating from the 3$d$ spins of $Fe^{3+}$ in T-$FeS_2$ monolayer.

In order to understand the origin of ferromagnetism observed in few-layer 1T-$FeS_2$ crystals, the density functional theory (DFT) method was employed with advanced SCAN+rVV10 exchange-correlation functional.[45, 46, 47] The experimental lattice ($a$ = 3.36 Å) was adopted to relax the atomic structure. Interlayer spacing is evaluated to 3.04 Å for 3D layered structure, while it enlarges to 3.23 Å for bilayer and 3.30 Å for trilayer. To determine the ground magnetic ordering of monolayer, a 2$\sqrt{3}$a×2a orthorhombic cell was built with ferromagnetic (FM), stripe antiferromagnetic (sAFM) and two zigzag antiferromagnetic (zAFM1 and zAFM2) ordering shown in Supplementary Fig. 12. Clearly, the ferromagnetic ordering has the lowest energy, confirming it is the ground ferromagnetic state, while the antiferromagnetic ordering configurations have larger energy by 0.26 eV (sAFM), 0.24 eV (zAFM1) and 0.53 eV (zAFM2), referring to FM ordering. Also, the energy of the interlayer ferromagnetic configuration of bilayer is lower than that of interlayer antiferromagnetic configuration by 2.58 meV/cell. Therefore, the ferromagnetic coupling is the ground state for layered 1T-$FeS_2$. As shown in Fig. 5a,b, 1T-$FeS_2$ monolayer presents a half-metallicity highlighted by fully spin-polarization around Fermi level, of which the band structure is projected orbitally. Clearly, the ferromagnetism is primarily attributed to the splitting of Fe $d$ orbitals, while the S $p$ orbitals participate in the magnetic coupling around Γ point near Fermi level. Similarly, the half-metallicity is also presented in bilayer and trilayer, the band structure of which is shown in Supplementary Fig. 13. Alternatively, the ferromagnetic coupling of 1T-$FeS_2$ can be evaluated by Stoner criterion (*ST*). If *ST* > 1, a metallic system can be defined to have a ferromagnetic ground state in principle (see Supplementary Note 1 for more details). Through the spin-polarized band structure of 3D 1T-FeS$_2$ in Supplementary Fig. 14, we could get *ST* = 1.30 for few layer 1T-FeS$_2$, suggesting that 1T-FeS$_2$ has a stable Stoner ferromagnetism.

To demonstrate the magnetic anisotropy that exists in 1T-FeS$_2$, we introduce the magnetic anisotropy energy with the form Δ*E* = *E*$_\perp$ - *E*$_\parallel$, where *E*$_\perp$ and *E*$_\parallel$ represent the energy of the configuration with out-of-plane magnetic moments and in-plane magnetic moments, respectively. The magnetic anisotropy of monolayer, bilayer, trilayer and bulk is estimated to be -0.12 eV/Fe, -0.06 eV/Fe, -0.01 eV/Fe and -0.04 eV/Fe, respectively. Apparently, monolayer has the strongest magnetic anisotropy with out-of-plane ferromagnetism. The variation of the magnetic anisotropy in bilayer and trilayer is dominated by the interlayer interaction reflected by the interlayer spacing. These all indicate that 1T-FeS$_2$ shows out-of-plane ferromagnetism with Ising highlighted, which is in line with our experimental observation.

Considering few defects and vacancies in 1T-FeS$_2$ crystal as observed in STEM characterization, we also studied their effects on the emergent magnetism. As shown in Fig. 5c, the introduction of a few sulfur vacancies will not induce obviously structural distortion of the 1T-FeS$_2$ structure and it is similar to the experimental observation, in which the irons are still surrounded by spin-polarized *d* electron density. Figure 5d shows the spin-polarization of atom-projected *d* electron density of states of pristine and defect system, defined as ($\rho\uparrow$-$\rho\downarrow$)/($\rho\uparrow$+$\rho\downarrow$) where $\rho\uparrow$ and $\rho\downarrow$ indicate the spin-majority and spin-minority electron density of states. The spin polarization of pristine 1T-FeS$_2$ reaches 1 around Fermi level. However, the defect system shows a reduction of spin-polarization, indicating that the sulfur vacancy provides an impact on the magnetic coupling in 1T-FeS$_2$. With the further analysis of magnetic moment of iron atom in defect system (Fig. 5e), the sulfur vacancy induces an uneven ferromagnetic order, as manifested by the increase of magnetic moment of three nearest neighbor irons around the vacancy. Such an additional contribution of magnetic moment induced by sulfur vacancy is probably to form some novel spin texture, skyrmions for instance.

In summary, we have demonstrated the successful synthesis of theoretically predicated $FeS_2$ few layers—1T-$FeS_2$, a new kind of 2D ferromagnetic vdW material, via a facile CVD method. The intrinsic long-range 2D ferromagnetic order has been established by the observations of a temperature-dependent anomalous Hall effect (AHE) and a hysteretic magnetoresistance. The out-of-plane ferromagnetism is consistent with the first-principle calculations and the impact of sulfur vacancy is also discussed. Our findings on the 2D metallic ferromagnet of 1T-$FeS_2$ are expected to enrich the family of 2D magnetic materials, and boost the possible applications in spintronic devices, such as spin filters and spin valves.

## Methods

**Controllable synthesis of few-layer 1T-$FeS_2$ crystals.** High-purity elemental Sulfur and Iron Chloride ($FeCl_2$) from (99.99% Sigma Aldrich) are used for the synthesis of 1T-$FeS_2$ few layers. Alumina crucibles containing the precursors were loaded in a 1-inch quartz tube furnace. 285 nm Si/$SiO_2$ substrates were placed near the iron precursor and Argon gas (60 sccm) was used as the carrier gas. The $FeCl_2$:KCl ratio was maintained at 1:2 for the optimum growth. Thermcraft furnace Protégé 1100 °C Split Tube Furnace was used for the reaction. The furnace was heated to a temperature region between 600-650 °C for the growth at a ramp rate of 50 °C/min. The holding time was varied between 2-5 minutes for an optimal growth. The furnace was cooled down rapidly to avoid thicker $FeS_2$ flakes. The layer number increases with the growth temperature. Monolayer $FeS_2$ crystals were mainly obtained when $T$ was set to 600–620 °C. The optimum Fe:KCl ratio for few-layer FeS$_2$ synthesis was 1:2 (Supplementary Fig. 5). The distance between the S source and Fe Source is found to be ~ 10 cm for the optimum growth of FeS$_2$ flakes (Supplementary Fig. 6).

**Sample characterizations.** STEM imaging and EELS experiments were performed on a JEOL ARM200 microscope equipped with a Schottky field emission gun, a JEOL double Wien filter monochromator, probe delta corrector and a Gatan Quantum GIF spectrometer modified for low primary energy operation (15-60 keV) with high stability. Samples for TEM investigations were prepared by adding ethanol to the samples onto a graphene substrate supported on a copper grid. The excess liquid was then removed, and the grid was allowed to dry at room temperature. XPS spectra were collected on a SPECS I3500 plus spectrometer using Mg X-ray source. AFM images were taken using the Asylum Research Cypher AFM in tapping mode. Raman spectra were recorded in vacuum by a Witec system with ×50 objective lens and a 2400 lines per mm grating under a 532 nm laser excitation. The laser power was fixed at 1 mW.

**Device fabrication and transport measurement.** 1T-FeS$_2$ flakes were deposited on SiO$_2$/Si substrate, which facilitate the device fabrication without the need for transferring the materials to an insulating substrate for transport measurement. After the growth of the sample, 1T-FeS$_2$ flakes with the thickness ranging from 5 nm to 30 nm were firstly identified by their color contrast under optical microscopy. Then small markers were fabricated using standard e-beam lithography (EBL) near the identified sample for subsequent fabrication of Hall-bar devices. To obtain a clean interface between the electrodes and the sample, *in situ* argon plasma was employed to remove the resist residues before metal evaporation without breaking the vacuum. Subsequently, 5/80 nm-thick Ti/Au electrodes were deposited using an electron-beam evaporator, followed by a lift-off step. Low-temperature magnetotransport experiments were carried out with a standard four-terminal method from room temperature to 0.3 K in a top-loading Helium-3 refrigerator with a 15

T superconducting magnet. A standard low-frequency lock-in technique was used to measure the resistance with an excitation current of 1 μA.

**Computational method.** Density functional theory calculations were carried out using Vienna ab initio simulation package (VASP) with the implementation of the projector-augmented wave (PAW) pseudopotential and plane-wave basis set.[46, 48] The strongly constrained and appropriately normed (SCAN) meta-generalized gradient approximation (METAGGA) exchange-correlation potential was used to self-consistently solve Kohn-sham equations.[49] SCAN+rVV10 was chosen for the consideration of the weak van der Waals interaction in layered 1T-$FeS_2$.[50] In structure relaxation and static self-consistent calculations, Gamma-centered k-mesh of 15×15×1, 15×15×2 and 3×3×1 was set for few-layer, bulk and defect system, respectively. The plane-wave kinetic cutoff energy of crystals and defect system is set to 800 eV and 500 eV. Structure relaxation converged until the absolute force on each atom is less than 0.01 eV/Å. The spin-orbital coupling effect was considered for determining the magnetic ordering and the magnetic anisotropy. The code for post-analysis is available from corresponding author upon reasonable request.

**Acknowledgements**

The research at IOP is supported by the National Natural Science Foundation of China under grant numbers 92065203, 11527806, 61974120, and 11874406, by the National Basic Research Program of China from the MOST under grant numbers 2016YFA0300601, by the Beijing Municipal Science & Technology Commission of China under grant number Z191100007219008, and by the Strategic Priority Research Program of the Chinese Academy of Sciences under grant number XDB33010300. This work is supported by the Synergic Extreme Condition User Facility. This work was supported by National Research Foundation–Competitive Research Program NRF-CRP22-2019-0007 and NRF-CRP22-2019-0004). This work was also supported from the Singapore Ministry of Education Tier 3 Programme "Geometrical Quantum Materials" AcRF Tier 3 (MOE2018-T3-1-002), AcRF Tier 2 (MOE2019-T2-2-105), AcRF Tier 1 (RG7/18 and RG161/19).

**Author contribution**

GKR, XJ, DG and CZ contributed equally to this work. ZL and GL conceived and supervised the project; GKR synthesized and characterized the sample. XJ and DG fabricated the devices and carried out the low-temperature magnetotransport measurements; GKR and ZC did the measurements and data analysis on STEM; XZ, XY, JC, PL, XW, WY, TS, JZ, JY, FM, CY, HK, FQ, JS, XJ, and LLcontributed to the results analysis and discussions. XX and XL performed first-principles calculations. GKR, XL, ZL, and GL co-wrote the paper. All the authors discussed the results and commented on the manuscript.

**Additional information**

Supplementary information is available in the online version of the paper. Reprints and permissions information is available online at www.nature.com/reprints. Correspondence and requests for materials should be addressed to GL, XL and ZL.

 **Competing financial interests**

The authors declare no competing financial interests.

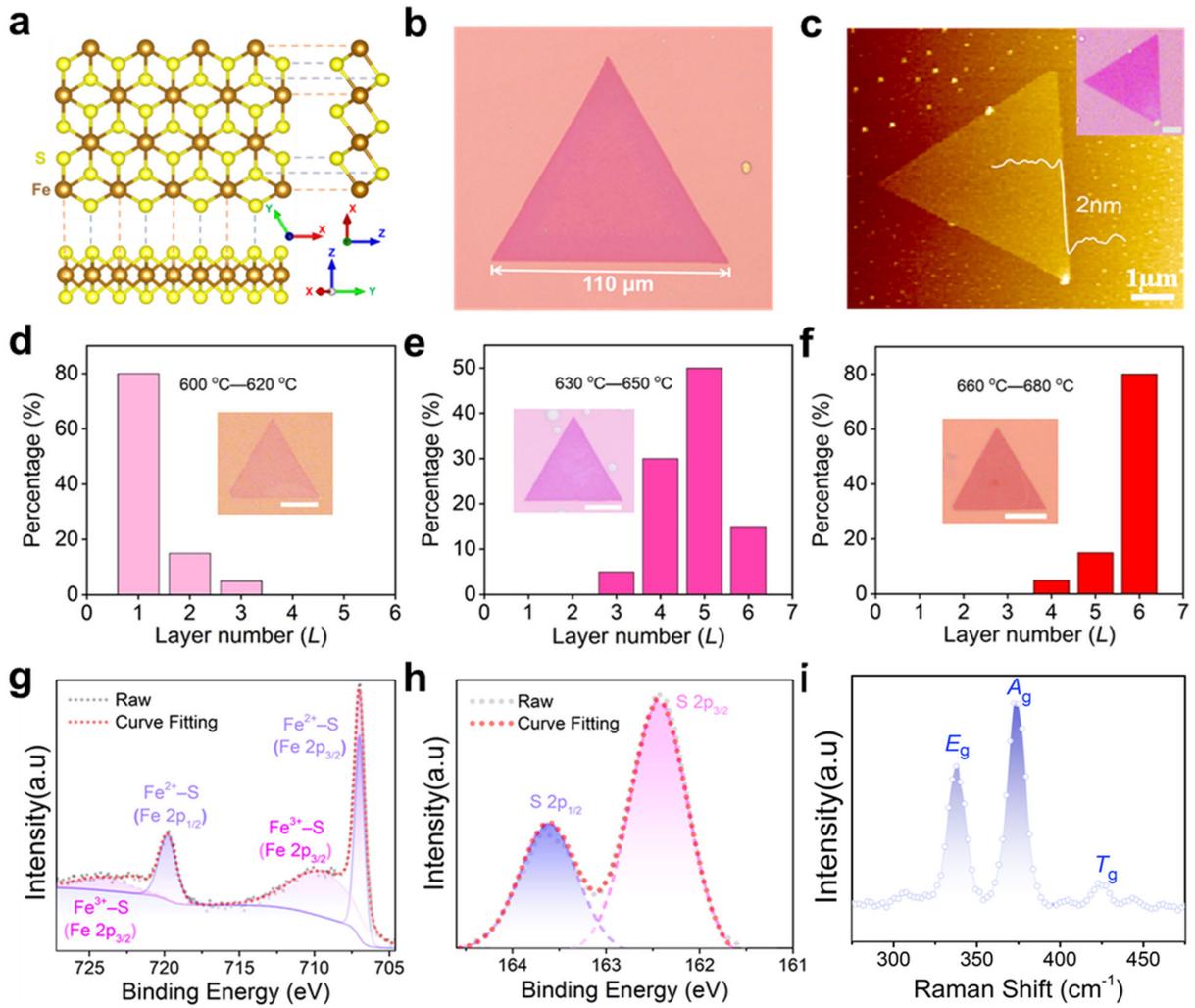

**Fig. 1 │ Atomic structure, morphologies, and characterizations of few-layer 1T-FeS$_2$ crystals. a**, Ball-and-stick model of monolayer 1T-FeS$_2$ viewed from three different crystallographic directions. **b**, Optical image of a large-area uniform 1T-FeS$_2$ flake deposited on a SiO$_2$/Si substrate with the edge length up to 110 μm. **c**, Atomic force microscopy (AFM) image of a typical atomically thin 1T-FeS$_2$ having a thickness of approximately 2 nm, indicating the sample is bilayer. Inset: the optical image of the sample used for AFM measurement. **d-f** Statistic thickness distributions and representative morphologies (inset) of FeS$_2$ crystals prepared at 600-620 °C (**d**), 630-650 °C (**e**) and 660-680 °C (**f**), respectively. Scale bars from inset of **d-f** are 5 μm. Thickness of inset crystals of **d-f** are referenced in the Supplementary Fig. 6. **g,h**, X-ray

photoemission spectroscopy (XPS) spectra of the Fe $2p$ (**g**) and S $2p$ (**h**) peaks from FeS$_2$ crystals deposited on SiO$_2$/Si substrate. **i**, Raman spectra of a typical 1T-FeS$_2$ flake.

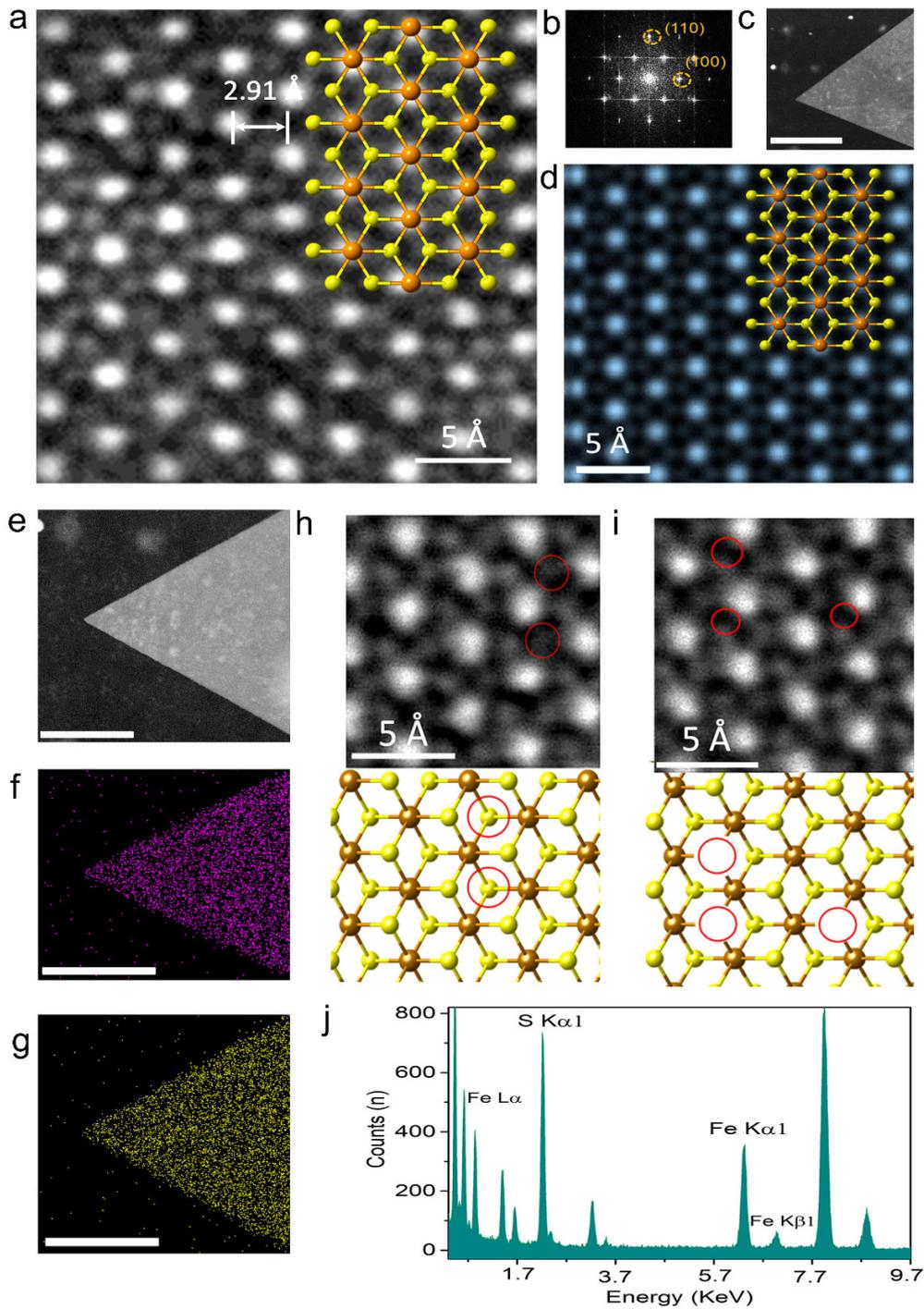

**Fig. 2 | Structural and elemental characterizations of as-grown 1T-FeS$_2$ few layers. a**, An atomic-resolution annular dark-field scanning transmission electron microscope (ADF-STEM) image of a triangular 1T-FeS$_2$ flake in the (001) direction. **b**, FFT pattern of the 1T FeS$_2$ with

crystal orientation planes (110) and (100) marked. **c**, low resolution STEM image of the 1T FeS$_2$ flake used for the imaging in (**a**). Scale bar is 500 nm. **d**, structural model are overlaid on the image, showing a perfect agreement with the experimental result. **e**, STEM image of a large region of FeS$_2$ used for the collection of energy-dispersive X-ray spectrometry (EDX) spectra. **f,g,** Elemental mapping of Fe and S are provided next to it. Scale bar: 250 μm. **h,I,** Point lattice defects (top) and Sulfur (bottom) vacancies observed in the FeS$_2$ lattice along with the simulation model portraying the same. **j**, EDX spectra of the region shown in (**e**).

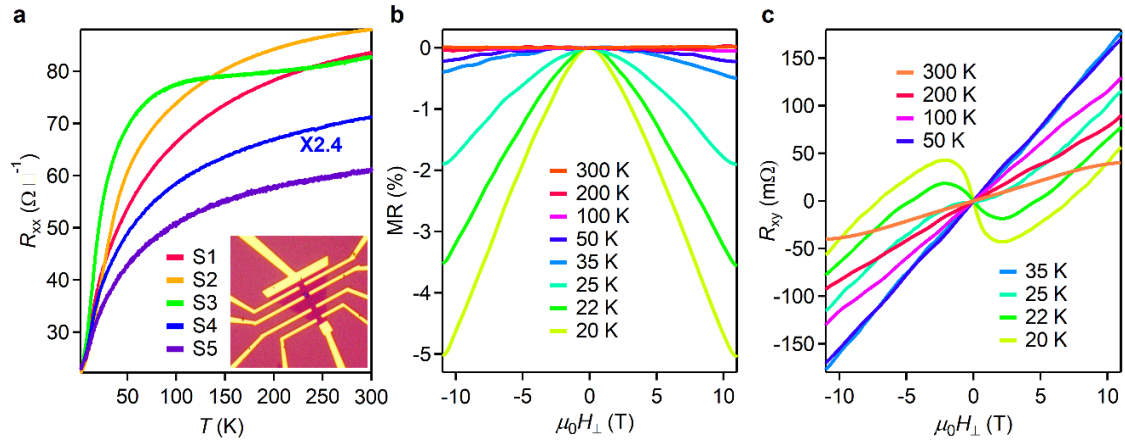

**Fig. 3 | Electrical transport properties for few-layer 1T-FeS$_2$ samples. a,** Temperature dependence of zero-field sheet resistance in five different ultra-thin 1T-FeS$_2$ devices (S1~S5). The sheet resistance curve of device S4 is multiplied by 2.4 for clarity. The inset shows a typical Hall-bar device used in low-temperature magnetotransport measurements. **b,c,** Temperature dependence of longitudinal magnetoresistance (**b**) and Hall resistance $R_{xy}(H)$ (**c**) for device S1 measured at different temperatures under perpendicular magnetic fields.

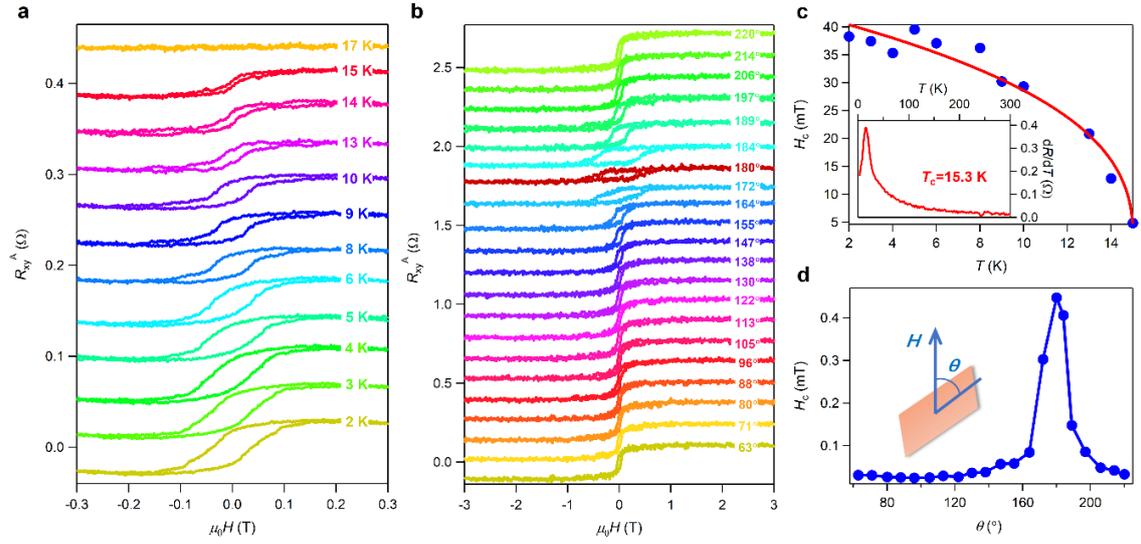

**Fig. 4 | Ferromagnetism realized in ultra-thin 1T-FeS$_2$. a**, Anomalous Hall effect (AHE) observed in device S4 at different temperatures. The coercive field and remanent Hall resistance $R_{xy}$ decreases as the temperature increases, until both vanish at 17 K. **b**, Tilt angle dependence of the AHE observed at $T$=0.3 K in device S5. The curves in (**a**) and (**b**) are shifted vertically for clarity and easier comparison. **c**, Temperature dependence of the coercive field $H_c(T)$ extracted from (**a**) for device S4. The red dashed line is a fit with an emprical formula $H_c \propto (1 - T/T_c)^\beta$. The inset shows the first derivative of the zero-field $R_{xx}(T)$ as a function of temperature for device S4, where the peak temperature indicates the Curie temperature $T_c$=15.3 K. **d**, Angular dependence of the coercive field $H_c(T)$ for device S5. The inset is a schematic drawing of the tilt experimental setup, where $\theta$ is the out-of-plane tilt angle between the out-of-plane magnetic field $H$ and the sample surface.

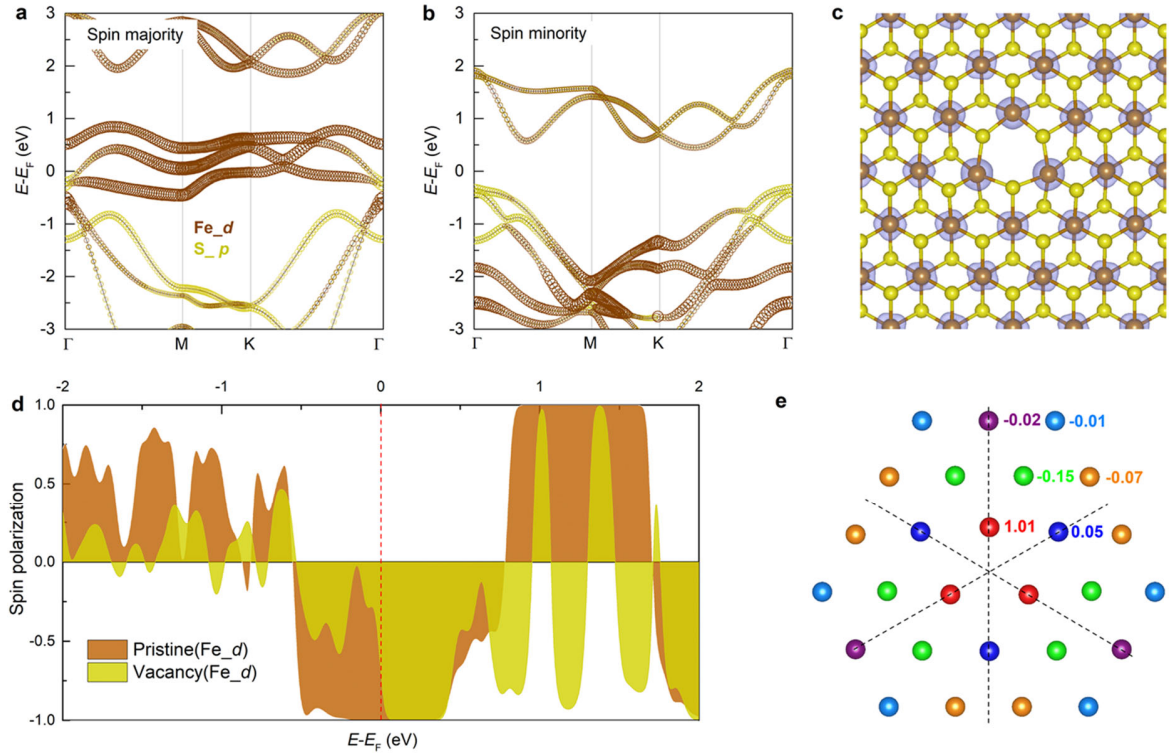

**Fig. 5 | The ferromagnetism of 1T-FeS$_2$ and sulfur vacancy impact. a,b,** the element-projected spin-majority (**a**) and spin-minority (**b**) band structure of monolayer 1T-FeS$_2$. The brown- and yellow-colored circle bands with the same scaling factor indicate the contribution from the iron *d* orbitals and sulfur *p* orbitals, respectively. **c,** The atomic structure of defect system with single sulfur atom vacancy, in which iron atoms are surrounded by spin-polarized *d* electron density with the isosurface value of 0.01. **d,** The spin-polarization of the density of states of pristine and defect 1T-FeS$_2$. The spin-polarization of defect system is projected on the red-colored iron atom in (**e**). **e,** The variation of magnetic moment on iron atoms induced by sulfur vacancy. The value alongside atoms is the change of magnetic moment referring to the magnetic moment (1.91 $\mu_B$) of iron in pristine system. The dash lines indicate the symmetry axis. The atoms with the same color are symmetrically equivalent.